\documentclass[aps, reprint, superscriptaddress,amsmath,amssymb]{revtex4-1}

\newcommand{\textunderscript}[1]{$_{\text{#1}}$}
\usepackage{float}
\usepackage{longtable}
\usepackage{supertabular} 
\usepackage{epstopdf}
\usepackage{array}
\usepackage{graphicx}
\usepackage{dcolumn}
\usepackage{bm}
\usepackage{hyperref}

\usepackage{lipsum}

\begin{document}
\preprint{APS/123-QED}


\title{Helicity-dependent all-optical domain wall motion in ferromagnetic thin films}



\author{Y. Quessab}
\email[]{yassine.quessab@univ-lorraine.fr}
\affiliation{Institut Jean Lamour, UMR CNRS 7198, Universit\'e de Lorraine, BP 70239, F-54506, Vandoeuvre-l\`es-Nancy, France}
\affiliation{Center for Memory and Recording Research, University of California San Diego, La Jolla, California 92093-0401, USA}

\author{R. Medapalli}
\affiliation{Center for Memory and Recording Research, University of California San Diego, La Jolla, California 92093-0401, USA}

\author{M. S. El Hadri}
\affiliation{Institut Jean Lamour, UMR CNRS 7198, Universit\'e de Lorraine, BP 70239, F-54506, Vandoeuvre-l\`es-Nancy, France}

\author{M. Hehn}
\affiliation{Institut Jean Lamour, UMR CNRS 7198, Universit\'e de Lorraine, BP 70239, F-54506, Vandoeuvre-l\`es-Nancy, France}

\author{G. Malinowski}
\affiliation{Institut Jean Lamour, UMR CNRS 7198, Universit\'e de Lorraine, BP 70239, F-54506, Vandoeuvre-l\`es-Nancy, France}

\author{E. E. Fullerton}
\affiliation{Center for Memory and Recording Research, University of California San Diego, La Jolla, California 92093-0401, USA}

\author{S. Mangin}
\affiliation{Institut Jean Lamour, UMR CNRS 7198, Universit\'e de Lorraine, BP 70239, F-54506, Vandoeuvre-l\`es-Nancy, France}


\date{\today}

\begin{abstract}
Domain wall displacement in Co/Pt thin films induced by not only fs- but also ps-laser pulses is demonstrated using time-resolved magneto-optical Faraday imaging. We evidence multi-pulse helicity-dependent laser-induced domain wall motion in all-optical switchable Co/Pt multilayers with a laser energy below the switching threshold. Domain wall displacement of $\sim$ 2 nm per 2-ps pulse is achieved. By investigating separately the effect of linear and circular polarization, we reveal that laser-induced domain wall motion results from a complex interplay between pinning, temperature gradient and helicity effect. Then, we explore the microscopic origin of the helicity effect acting on the domain wall. These experimental results enhance the understanding of the mechanism of all-optical switching in ultra-thin ferromagnetic films.
\end{abstract}

\maketitle


\section{Introduction}

Magnetization manipulation based on ultrashort laser pulses without any external magnetic field has recently attracted researchers' attention as it could lead to ultrafast and high-density magnetic data storage \cite{ElHadri2017, Kirilyuk2010}. In 2007, it was shown that the magnetization of a ferrimagnetic GdFeCo alloy could be fully reversed on a ps time scale using circularly polarized light \cite{Stanciu2007}. Thus, all-optical switching (AOS) rapidly became a topic of great interest. It was observed in a wider variety of materials ranging from ferrimagnetic multilayers and heterostructures, rare-earth (RE)-free synthetic ferrimagnetic heterostructures \cite{Mangin2014,Schubert2014} to ferromagnetic continuous thin films and granular media \cite{Lambert2014,Takahashi2016}. Unlike GdFeCo alloys for which all-optical magnetization reversal is said to result from a pure thermal process \cite{Radu2011, Ostler2012, Khorsand2012}, several mechanisms and microscopic models based on the Inverse-Faraday Effect (IFE) or Magnetic Circular Dichroism (MCD) were proposed to explain all-optical helicity-dependent magnetization switching (AO-HDS) in ferromagnetic materials \cite{Cornelissen2016, Gorchon2016}. Furthermore, single-pulse optical excitation of Pt/Co/Pt only leads to thermal demagnetization [13]. Thus, the latter indicates that in ferromagnetic systems, all-optical switching is rather a cumulative and multi-pulse mechanism \cite{Takahashi2016, ElHadri2016a, Medapalli2016} with two regimes: a demagnetization and a multi-domain state followed by a helicity-dependent remagnetization assumed to result from domain wall (DW) motion which depends upon the light helicity \cite{ElHadri2016a, Medapalli2016}.

In continuous magnetic media, a DW separates two magnetic domains of uniform and opposite magnetization. Domain walls with their high mobility are of great interest for low-power spintronic applications, such as racetrack memories \cite{Parkin2008} or logic devices \cite{Hayashi2008, Allwood2005}. Magnetic field-driven DW motion in ferromagnets with strong perpendicular anisotropy, such as ultra-thin Pt/Co/Pt films, has been extensively studied \cite{Metaxas2007}. Current-induced DW motion in ferromagnetic elements, via spin-transfer torque (STT), was also reported \cite{Malinowski2011, Miron2011}. Control of the DW by electric field \cite{Lahtinen2012}, voltage-induced strain \cite{Shepley2015}, thermal gradients either by injecting current \cite{Torrejon2012} or local heating \cite{Tetienne2014, Moretti2017, Schlickeiser2014} are other possibilities for manipulation of DW. In this article, we report deterministic motion of domain walls in Co/Pt multilayers that show AO-HDS, using circularly polarized laser pulses. We have investigated the DW displacement as a function of laser polarization, beam position and laser power. The results reveal that the physical mechanism differs from pure thermal gradient-driven DW motion. Instead, it arises from the balance of three contributions: pinning, heating and helicity. Inverse-Faraday effect and magnetic circular dichroism are two models explored to elucidate the effect of helicity on the DW. These findings can be key elements to explain magnetization reversal in AOS.

\section{Samples and experimental methods}

Two ferromagnetic (Co/Pt) multilayers were studied: Glass/Ta(5)/Pt(5)/[Co(0.4)/Pt(0.7)]\textunderscript{x3}/Pt(2) and Glass/Ta(5)/Pt(4.5)/Co(0.6)/Pt(0.7)/Pt(3.8) (in brackets thickness in nm). These thin films were both grown by DC magnetron sputtering. The bottom Ta/Pt bilayer allows a good adherence of the multilayer stack on the glass substrate and a (111) texture of the (Co/Pt) layers, which ensures perpendicular magnetic anisotropy (PMA) and high anisotropy field \cite{Bersweiler2016}. The top Pt layer prevents sample oxidation.

\begin{figure*}[ht]
\begin{center}
\scalebox{1}{\includegraphics[width=17cm, clip]{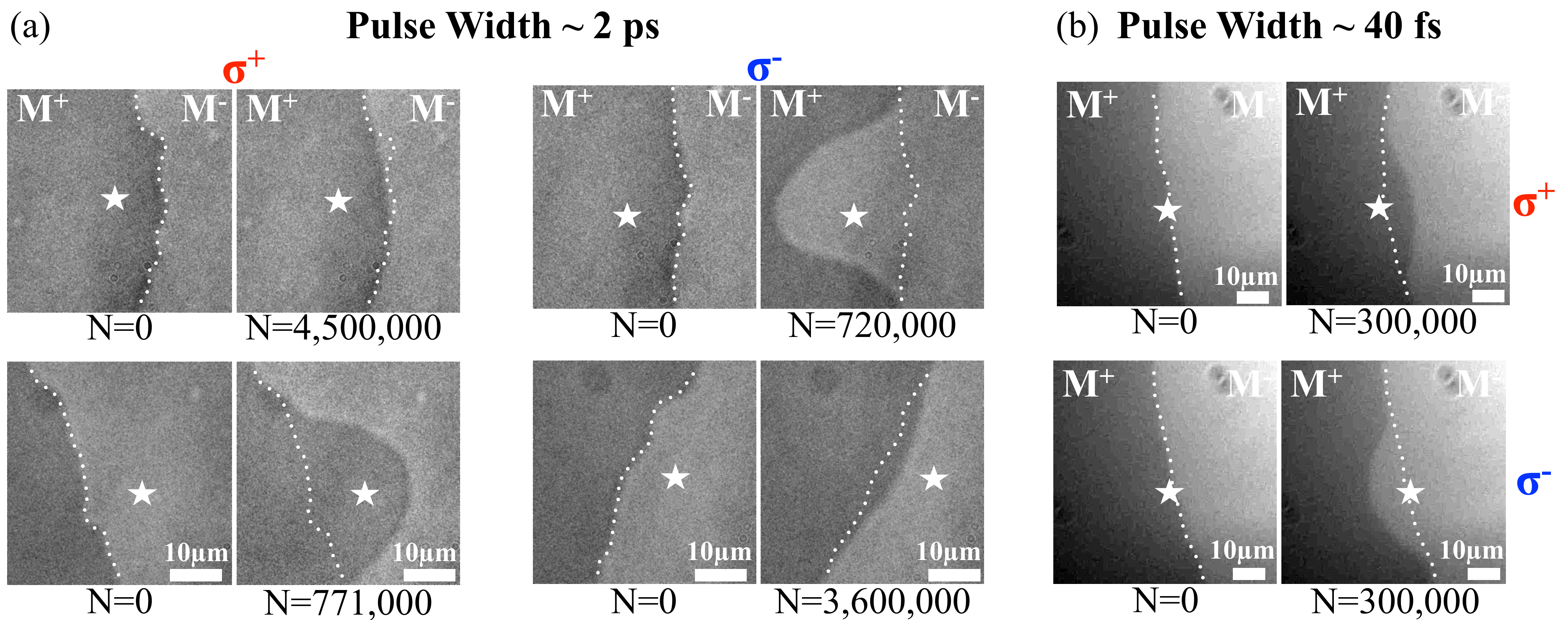}}
\end{center}
\caption{\label{sample}Magneto-optical images of domain wall motion in [Co(4{\AA}/Pt(7{\AA})]\textunderscript{x3} and [Co(6{\AA})/Pt(7{\AA})]\textunderscript{x1} induced respectively by (a) 2 ps- and (b) 40 fs- laser pulses with left- ($\sigma$\textsuperscript{-}) [right- ($\sigma$\textsuperscript{+})] circular polarization with an energy per pulse of 0.04 mJ.cm\textsuperscript{-2} and 12.5 mJ.cm\textsuperscript{-2}. The white star indicates the center of the beam spot and N the number of laser pulses. The laser beam spot is (a) placed 10 \textit{$\mu$}m away from the DW within a magnetization-up (\textit{M}\textsuperscript{+}) or -down (\textit{M}\textsuperscript{-}) domain or (b) centered on the wall. The dashed line shows the initial position of the domain wall prior to laser exposure.}
\end{figure*}

To perform optical excitation and AO-HDS, two different ultrafast laser systems were used. On one hand, the Pt/Co/Pt single layer sample was exposed to a Ti:Sapphire fs-laser with a 5 kHz repetition rate, a central wavelength of 800 nm and a pulse duration of 40 fs. The laser beam spot has a Gaussian profile and is focused onto the sample surface with a full-width at half maximum (FWHM) of 50 $\mu$m. On the other hand, Pt/[Co/Pt]\textunderscript{x3}/Pt  trilayer thin film was excited by 2-ps 800 nm laser pulses with a repetition rate of 1 kHz and a Gaussian beam spot with a FWHM of 45 $\mu$m. Right-circularly ($\sigma$\textsuperscript{+}), left-circularly ($\sigma$\textsuperscript{-}) or linearly ($\pi$) polarized light is obtained with the use of a polarizer combined with a quarter-wave-plate (QWP). These two ferromagnetic multilayer films were previously reported to exhibit AO-HDS respectively with the two aforedescribed laser systems \cite{ElHadri2016a, Medapalli2016}. To investigate laser-induced domain wall motion, we implemented a time-resolved magneto-optical Faraday imaging technique. To probe the effect of the optical excitation of the DW, a Faraday microscope was used and a CCD camera to take an image every second.\raggedbottom

Prior to the DW motion experiments, all-optical switching of the films was first verified by sweeping the laser beam over the sample surface. We determined the power threshold for which AO-HDS is observed. Afterwards, an external magnetic field perpendicular to the sample surface is applied to set the magnetization in the `up' (\textit{M}\textsuperscript{+}) direction which corresponds to dark contrast on Faraday images. Then, a reversed magnetic domain (\textit{M}\textsuperscript{-}, bright contrast) is created. During all DW motion experiments, no magnetic field is applied to the sample, and the laser pump comes at normal incidence on the sample surface. The center of the laser beam, i.e. the maximum of intensity, is placed at different positions with respect to the DW with a motorized micro-stage.\raggedbottom

\section{Results}
\subsection{Fs- and ps- laser-induced domain wall motion in Co/Pt thin films}

As already mentioned, it was recently demonstrated that Pt(4.5)/Co(0.6)/Pt(4.5) single layer and Pt(5)/[Co(0.4)/Pt(0.7)]\textunderscript{x3}/Pt(2) trilayer showed AO-HDS respectively with fs- and ps- laser pulses via electrical Hall measurements or static imaging after laser beam sweeping \cite{Lambert2014, Medapalli2016, ElHadri2016b}. First, we investigated whether we could observe DW motion in these materials induced solely by not only ps-laser pulses but also with even shorter fs-pulses (see Fig. 1). DW experiments were performed at a fluence lower than the switching threshold so that no reversed magnetic domain is observed. Therefore, the changes in the DW pattern in Fig. 1 can only be attributed to DW motion and not to domain nucleation. Fig. 1(a) shows the evolution of the DW in Pt/[Co/Pt]\textunderscript{x3}/Pt after being exposed to circularly polarized light. The center of the laser beam spot is placed at 10 $\mu$m from the DW either on a domain with a magnetization-up (\textit{M}\textsuperscript{+}) or magnetization-down (\textit{M}\textsuperscript{-}). Thus, four combinations of light polarization and position of laser beam ($\sigma$\textsuperscript{+}, \textit{M}\textsuperscript{+}), ($\sigma$\textsuperscript{+}, \textit{M}\textsuperscript{-}), ($\sigma$\textsuperscript{-}, \textit{M}\textsuperscript{+}) and ($\sigma$\textsuperscript{-}, \textit{M}\textsuperscript{-}) were studied. Note that the laser beam spot overlaps both magnetic domains, yet what is important is the position of the maximum of intensity, i.e. the location of the hottest region with regard to the DW.  When clear DW displacement (DWD) was observed, image recording was stopped after stabilization of the DW. The experiments were repeated several times, every time on a different area that was not previously exposed to the laser beam. Only two combinations, ($\sigma$\textsuperscript{+}, \textit{M}\textsuperscript{-}) and ($\sigma$\textsuperscript{-}, \textit{M}\textsuperscript{+}), led to significant DW displacement. As seen on Fig. 1(a), when a \textit{M}\textsuperscript{-} domain is exposed with $\sigma$\textsuperscript{+} pulses, the DW moves such as the \textit{M}\textsuperscript{+} domain expands. Conversely, a \textit{M}\textsuperscript{+} domain illuminated with $\sigma$\textsuperscript{-} polarization leads to an expansion of the \textit{M}\textsuperscript{-} domain. Note that in sweeping measurements, $\sigma$\textsuperscript{+} ($\sigma$\textsuperscript{-}) polarization reverses \textit{M}\textsuperscript{-} (\textit{M}\textsuperscript{+}). Same results were observed in Pt/Co/Pt single layer exposed to 40-fs laser pulses as depicted in Fig. 1(b). In this case, the laser beam spot is centered on the DW and  the displacement direction of the DW is determined by the helicity of the laser pulses. Later, we found that the laser-induced DW motion could be cancelled with an out-of-plane magnetic field of about 2 Oe whose direction depends on the light helicity that was used. Interestingly, applying 2 Oe perpendicularly to the sample was also sufficient to cancel AO-HDS when sweeping circularly polarized light. This field is in the same order of magnitude than what was previously reported for Co/Pt multilayer thin films by C.-H. Lambert \textit{et al}. \cite{Lambert2014}. Thus, it is very clear that the direction of the DW displacement depends on the light helicity in all-optical switchable Co/Pt systems, and that it corresponds to the reversal direction observed in AO-HDS. 

\begin{figure}[ht]
\begin{center}
\scalebox{1}{\includegraphics[width=7.5cm, clip]{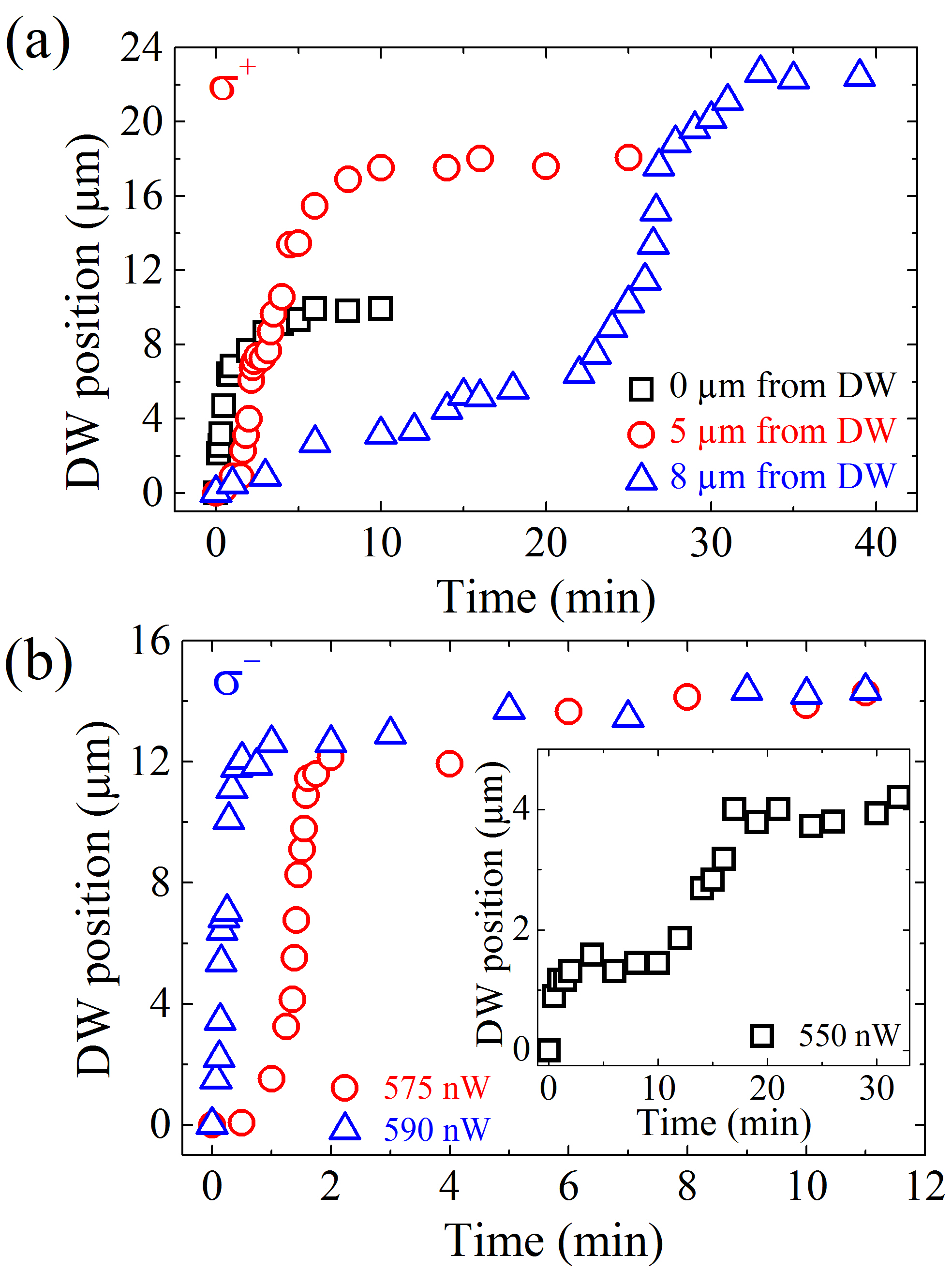}}
\end{center}
\caption{\label{sample}Time evolution of the domain wall displacement (DWD) in Co(4{\AA})/Pt(7{\AA}) trilayer obtained from magneto-optical image recording as a function (a) of laser beam position from the DW and (b) laser fluence, upon excitation of a 2-ps laser beam with a repetition rate of 1 kHz. (a) right-circularly polarized light ($\sigma$\textsuperscript{+}) is used to shine a magnetization-down domain with a fluence of 0.04  mJ.cm\textsuperscript{-2} for a laser beam placed at 0, 5 and 8 $\mu$m from the DW. (b) left-circularly polarized ($\sigma$\textsuperscript{-}) laser beam is placed at 5 $\mu$m from the DW on a domain with magnetization down with three different laser powers. 3 regimes can be distinguished for the laser-induced DW motion: a slow displacement and depinning of the DW followed by a dramatic change of the DW position that finally stabilizes.}
\end{figure}

Thereafter, the dynamics of the DW displacement was explored at longer time scales as a function of the distance DW - laser beam and laser fluence in Pt/[Co/Pt]\textunderscript{x3}/Pt as shown in Fig. 2(a) and 2(b). Time-resolved measurements of DW motion were carried only for the configurations resulting in significant and measurable DW displacement. In Fig. 2(a) right-circular ($\sigma$\textsuperscript{+}) polarization and 2-ps laser pulses were used to observe DW motion for different positions of the center of the laser spot with respect to the DW within a \textit{M}\textsuperscript{-} domain. First of all, in Fig. 2(a) one can see that the further the laser beam, the greater the final DW displacement. Secondly, these plots reveal that the DW motion can be decomposed in 3 distinct regimes: the DW slowly starts moving, then it experiences a rapid displacement as the DW gets closer to the center of the laser spot and finally the speed decreases and the DW reaches a stable position. The first regime is absent when the center of the beam is on the DW, while the 3 regimes are clearly observed when the beam is 5 or 8 $\mu$m away from the DW. Furthermore, taking the derivative of the time evolution of the DW position, the DW velocity profile is obtained and has the shape of a Gaussian distribution. It distinctly exhibits the 3 afore-discussed regimes. Independently of the laser beam position, the maximum velocity is constant and is about 20 $\mu$m.min\textsuperscript{-1}, which corresponds to a displacement of $\sim$ 0.3 nm per 2-ps pulse for relatively low fluences of 0.04 mJ.cm\textsuperscript{-2}. Note that we were limited by the time resolution of 1 second, i.e. 1000 pulses, to calculate the DW mobility. 

Moreover, the laser power dependence of the DW motion was studied as shown in Fig. 2(b). Left-circular ($\sigma$\textsuperscript{-}) polarization was used to induce DW motion and the beam spot was placed at 5 $\mu$m from the DW within a \textit{M}\textsuperscript{+} domain. As depicted in Fig. 2(b) the higher the power, the faster the DW motion and the larger the displacement. Increasing the laser power only from 550 nW to 590 nW leads to a dramatic increase of the maximum DW displacement from 4 $\mu$m to 14 $\mu$m and the peak velocity from 1 $\mu$m.min\textsuperscript{-1} to 100 $\mu$m.min\textsuperscript{-1}. The DW reaches the same final position for P = 575 nW and 590 nW, thus, indicating that there is another limiting factor in addition to the laser fluence that controls the maximum DW displacement. This factor is likely to be related to the presence of pinning sites and a distribution of pinning energy in the continuous film. A maximum DW displacement of $\sim$ 2 nm per 2-ps pulse was achieved for the highest fluence. Note that the power window in which significant DW motion is observed is extremely narrow. For a power larger than 590 nW, nucleation started to take place, on the contrary for a power below 550 nW no DW displacement was measured, similarly for a beam position $>$ 10 $\mu$m. This indicates the existence of a power threshold, and thus a maximum initial distance between the DW and the center of the laser, to achieve DW displacement. This maximum initial distance is deduced from the abscissa of the power threshold on the Gaussian laser profile. Hence, the study of the DW dynamics allows to correlate the depinning time to the energy brought by the laser and the beam position, i.e. the spatial energy (temperature) profile that the DW sees. However, such measurements with our experimental setup were not possible for fs-laser pulses-induced DW motion as the displacement takes over a time scale that is much shorter than the time resolution of the time-resolved Faraday imaging technique we implemented. 

\subsection{Effect of the helicity and linear polarization on the domain wall motion}

To understand how DW motion is induced by light, it is important to separate the effects due to the temperature, as the laser brings heat to the sample, but also to the helicity. For this reason we reproduced the same experiments described in Fig. 1 but with linear polarization ($\pi$) for which only an increase of the temperature is associated to the optical excitation. The results are shown in Fig. 3(a). In Pt/Co/Pt single layer, a DW was created and then shined with linearly polarized light for three different laser beam positions. The exposure time was greater than the one needed to observe DW motion in Fig. 1(b). For a laser beam spot centered on the DW, no motion is observed (Fig. 3(a)). This can be understood in the sense that the temperature profile with respect to the DW is symmetrical and, consequently, no specific direction for the DW to move is preferred. While, as seen in Figs. 3(b) and 3(c), when the laser beam is off-centered either on magnetization-up or -down domain, the DW moves towards the center of the laser beam. If the fluence is too low, no significant DW displacement is observed, thus indicating that the energy brought by the laser was not enough to overcome the pinning energy barrier. These findings clearly demonstrate that in the absence of helicity the DW tends to move towards the hottest spot. 

\begin{figure}[ht]
\begin{center}
\scalebox{1}{\includegraphics[width=7cm, clip]{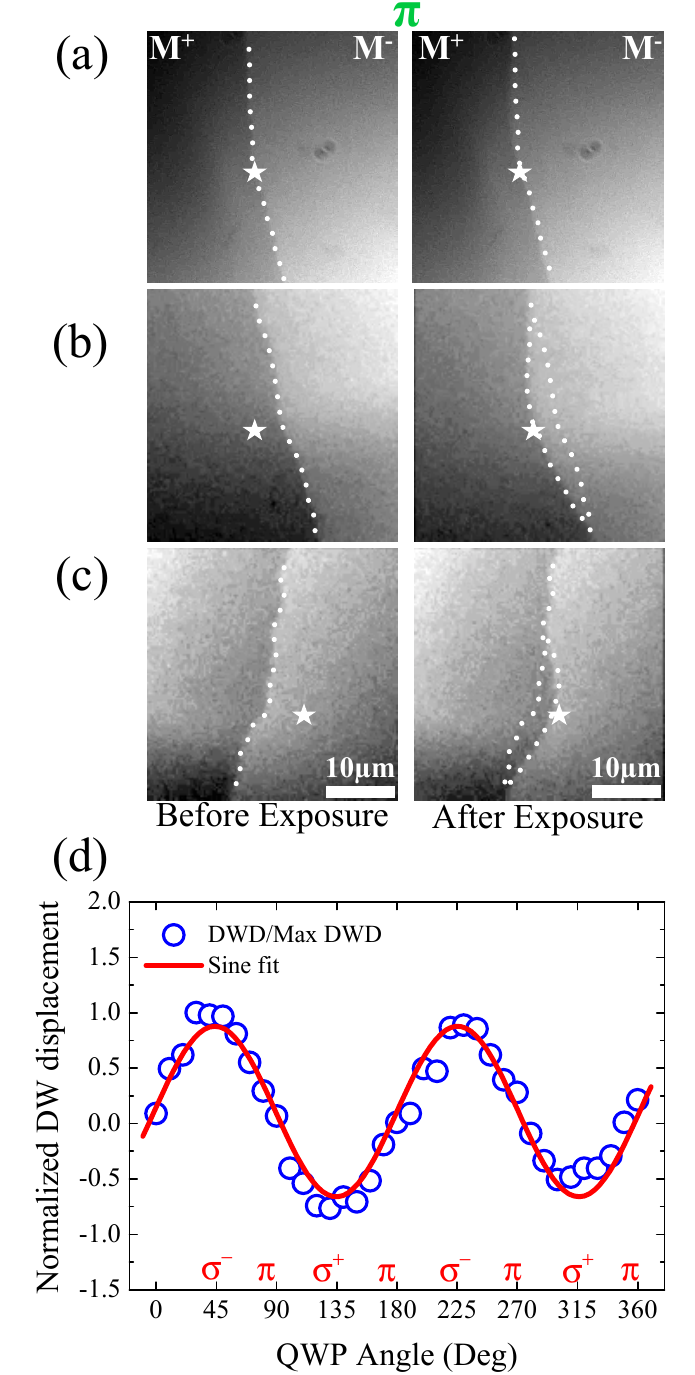}}
\end{center}
\caption{\label{sample}(a) and (b) Magneto-optical Faraday images of a domain wall (DW) in Pt(4.5nm)/Co(0.6nm)/Pt(4.5nm) thin film exposed to 40-fs linearly polarized ($\pi$) laser pulses with a fluence of 7 mJ.cm\textsuperscript{-2}. The laser beam spot (star) is (a) on the DW and off-centered in (b) and (c). The DW moves towards the center of the beam, i.e. the hottest regions, independently of the magnetization direction. (d) Normalized Domain wall displacement (DWD) induced by 40-fs laser pulses in Pt(4.5nm)/Co(0.6nm)/Pt(4.5nm) is plotted against the angle $\theta$ of the quarter-wave plate (QWP), i.e. the percentage of light helicity. The laser beam is initially centered on the DW at $\theta$ =  0$^{\circ}$ and kept fixed. The fluence is set to 12.5 mJ.cm\textsuperscript{-2}}
\end{figure}

Thereafter, we investigated the influence of the percentage of light helicity in the DWD and we measured the furthest stable DW position while gradually changing the angle $\theta$ of the quarter-wave-plate (QWP), i.e. progressively introducing or reducing helicity in the optical excitation (see Fig. 3(d)). A domain wall was created in the same material as previously. Initially, the center of the laser beam was placed on the DW and the angle of the QWP was set to 0$^{\circ}$. The polarization was changed by a step of 10$^{\circ}$ and an image with a Faraday microscope was taken only after stabilization of the DW. The results are presented in Fig. 3(d). The DWD is defined as the relative motion of the DW with respect to its initial position. The maximum DWD is reached when the sample is illuminated with circular polarization ($\sigma$\textsuperscript{+} and $\sigma$\textsuperscript{-}). Notably, the amplitude of the displacement is the same for both helicities, only the direction of the DW motion differs. 

Moreover, the position of the DW when excited by linear polarization ($\pi$) is close to 0, i.e. its initial position, which is in agreement with the findings in Fig. 3(a). Besides, the evolution of the DW position with the polarization can be fitted with a sinusoid, which indicates that the laser-induced DW motion is perfectly reproducible and robust with regard to the polarization that is used. Fig. 3(d) clearly shows that, although, circular and linear polarizations bring the same photon energy to the system, the DW ends up in a different potential well. Indeed, as seen in Fig. 3(b) and Fig. 3(c) linear polarization brings the DW towards the hottest spot, while circularly polarized light tends to move the DW towards colder regions, i.e. away from the center of the beam. However, even circularly polarized light brings heat to the sample, therefore the resulting displacement must be seen as a balance between the effect of helicity and temperature increase. Indeed, as the percentage of helicity increases, e.g. for $\theta$ going from 0 to $\sigma$\textsuperscript{-}, the relative displacement of the DW is greater, but from $\theta$ = $\sigma$\textsuperscript{-} to $\pi$ the effect of the helicity becomes weaker and the temperature gradient is in comparison higher, which tends to bring the DW back to its initial position. Thus, these experiments gives a clear evidence that helicity-dependent all-optical DW motion triggered by laser excitation results from the balance of 3 contributions, namely, the DW pinning, the effect of the light helicity and the temperature gradient induced by laser heating.

\section{Discussion}

The main results of this study demonstrate that it is possible to have helicity-dependent laser-induced DW motion in Co/Pt multilayer thin films. This corroborates the assumption made for the cumulative and two regimes switching process proposed to explain AO-HDS in ferromagnetic Co/Pt films \cite{ElHadri2016a, Medapalli2016}. Indeed, starting from a multi-domain state, it is clear now that under optical excitation with circularly polarized laser pulses, the DW will move in one direction according to the light helicity, which will result in the shrinkage or growth of domains of opposite magnetization. It is, now, important to understand the mechanism behind this helicity-dependent laser-induced domain wall motion in Co/Pt.

\subsection{Mechanism for laser-induced domain wall motion}

The experimental results in this article allowed us to exhibit three contributions that were involved in the DW displacement in Co/Pt induced by fs- and ps-laser pulses: DW pinning, temperature gradient across the DW and the effect of the helicity. To unpin a DW, an energy barrier \textit{E}\textunderscript{dep} has to be overcome associated with a depinning field \textit{H}\textunderscript{dep}. When the laser fluence was too low, no DWD was observed. Thus, indicating that the pinning potential was not overcome. As a result, pinning opposes to the laser-induced displacement direction. Moreover, exposing the sample to laser pulses generates heating in the material and, therefore, a temperature gradient across the DW. The influence of the increase of the temperature can be elucidated by studying the effect of linear polarization. Our results prove that the DW moves towards the hottest spot, which is consistent with previous studies that reported DW motion in thermal gradients in ferromagnetic systems \cite{Tetienne2014, Moretti2017, Schlickeiser2014}. Hence, the direction of the DWD is given by the direction of the temperature gradient. In the presence of a symmetric temperature gradient across the wall, e.g. when the laser is centered on the DW, no displacement is observed. However, regarding the effect of circular polarization upon the DW is more ambiguous as $\sigma$\textsuperscript{+} and $\sigma$\textsuperscript{-} polarizations also bring heat to the sample in addition to angular momentum. Yet, pure helicity effect can be revealed for a laser spot centered on the DW since in this configuration heating cannot break the symmetry  (see Fig. 1(b)). In this case, the direction of the displacement induced by pure helicity effect can be determined. 

Let's take the example of $\sigma$\textsuperscript{+} polarization as seen in Fig. 1(a). When the center of the laser spot is on a \textit{M}\textsuperscript{+} domain, the temperature gradient tends to pull the DW towards the hottest area, i.e. to the left, and the helicity in the other direction. The pinning aims to maintain the DW at the same position. As a result, all the effects cancel each other out and no significant displacement is observed.  When $\sigma$\textsuperscript{+} illuminates a \textit{M}\textsuperscript{-} domain, the temperature gradient and the helicity add up and are stronger than the pinning. Thus, they both pull the DW to the same direction. Once the DW crosses the center of the beam spot, the temperature gradient changes direction, and as it keeps moving further, the temperature gradient and the pinning get stronger and tend to compete with the helicity. Hence, DW motion continues until equilibrium of the three contributions is reached. This complex interplay between pinning, temperature and helicity also appears in Fig. 3(d). Since the laser is kept at a fixed position (normalized DWD = 0 in Fig. 3(b)), it is clear that any amount of helicity introduced in the light balances the temperature gradient and the pinning. 

\subsection{Models for the effect of the helicity on the domain wall}

In this section, we will discuss the microscopic origin of the effect of the helicity and its contribution to the laser-induced DW motion. Several mechanisms explaining AO-HDS in ferromagnetic thin films can be found in the literature based on either inverse Faraday effect (IFE) \cite{Cornelissen2016} or magnetic circular dichroism (MCD) \cite{Gorchon2016, Khorsand2012}. Here, we discuss two hypotheses to explain light-induced DW motion based either on athermal or pure thermal effects. First, we decided to test the IFE using the Fatuzzo-Labrune model that allows calculating the DW velocity in the case of magnetization relaxation (see Eq. 1) \cite{Fatuzzo1962, Labrune1989}. The model describes the energy that has to be brought by applying a magnetic field to overcome the pinning and make the DW move within a given volume, the Barkhausen volume (\textit{V}\textunderscript{B}). Moreover, the laser beam can be described as a Gaussian distribution of temperature and positive effective magnetic field for $\sigma$\textsuperscript{+} polarization under the assumption of IFE as shown in Fig. 4(a). We assume that the center of the laser beam spot corresponds to a maximum temperature of 600 K (300 K above room temperature) close to the Co Curie temperature and a field of 10 mT. IFE-induced field is said in the literature to vary from 0.1 to 30 mT \cite{Cornelissen2016}. Note that the tested field differs from the small out-of-plane field used to cancel the DW motion and the AOS. Indeed, the latter is permanent, while the IFE-induced field has a lifetime in the ps timescale \cite{Cornelissen2016}. 

\begin{equation}
\label{delta_H}
\  v_{H,T} = v_0  \exp\left(-\frac{E_{dep}-2HM_{S}V_{B}}{k_{B}T} \right)  
\end{equation}

This description of the laser beam is implemented in Eq. 1, which gives the DW velocity profile in Fig. 4(a). Each combination of temperature and effective magnetic field (\textit{T}, \textit{H}) at a distance x from the center of the beam generates a displacement u\textunderscript{\textit{T},\textit{H}(x)} at velocity v\textunderscript{\textit{T},\textit{H}(x)} calculated from the Fatuzzo-Labrune model. Assuming that the DW velocity is constant over $\delta$x$\ll$1, one can deduce v(t) and integrate it to obtain the time evolution of the DW position presented in Fig. 4(b) and (c) for the following tested values \textit{E}\textunderscript{dep} $\simeq$ 65 kbT, \textit{V}\textunderscript{B} = 9.7*10\textsuperscript{-18} cm\textsuperscript{-3}, \textit{v}\textunderscript{0} = 1.50*10\textsuperscript{-7} m.s\textsuperscript{-1}. The model was run for several laser beam positions (0, 5 or 8 $\mu$m from the DW). The simulations in Fig. 4(b) provide similar DW dynamics and results than previously described in Figs 2. Yet, it is noticeable that for a laser beam at 8 $\mu$m from the DW, the simulated displacement takes over a shorter timescale than in Fig. 2(a). This can be explained by the fact that the developed model does not take into account the pulse duration and the laser repetition rate, which are likely to impact the DW dynamics. Note that the simulations are not obtained by fitting our data. Implementing a left-circularly polarized laser ($\sigma$\textsuperscript{-}), i.e. a distribution of negative effective magnetic field leads to a vanishing velocity profile. As a result, helicity-dependent laser-induced DW motion in Pt(4.5nm)/Co(0.6nm)/Pt(4.5nm) was successfully reproduced. Thus, this demonstrates that the effect of the helicity on the DW can, indeed, be described as an athermal effective magnetic field whose direction depends upon the helicity, and that the DW displacement arising from this model is in agreement with our experimental data.

\begin{figure}[ht]
\begin{center}
\scalebox{1}{\includegraphics[width=7.5cm, clip]{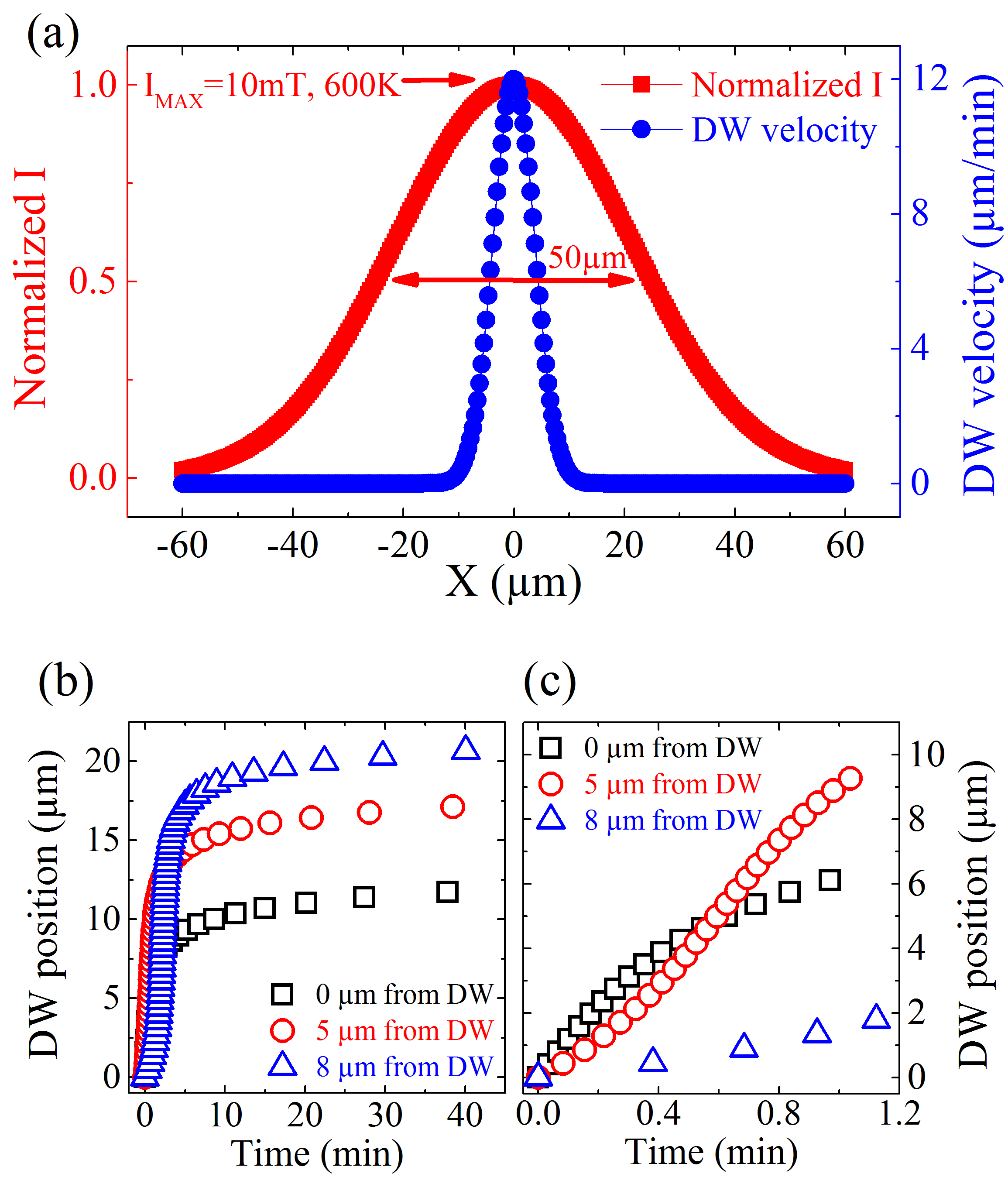}}
\end{center}
\caption{\label{sample}Simulations of the domain wall (DW) displacement in a Co(6{\AA})/Pt(7{\AA}) single layer thin film induced by circularly-polarized laser pulses (a) modeled as a Gaussian distribution of effective magnetic field and temperature with a maximum intensity of (10 mT, 600 K) and a full width at half maximum of $\sim$ 50 $\mu$m. (b) and (c) time evolution of the DW motion for three different distances between the laser beam and the DW based on the Fatuzzo-Labrune model. Similar results for the DW motion as described in FIG. 2. are obtained here.}
\end{figure}

Lastly, one could argue that the DW motion results only from pure thermal effects, namely MCD in addition to the DC laser heating close to Curie temperature. The difference in absorption of \textit{M}\textsuperscript{+} and \textit{M}\textsuperscript{-} domains, since in our experiments the laser beam spot overlaps with two domains of opposite magnetization directions, would result in an additional temperature gradient across the DW. As already mentioned, it was proven that DW could effectively move in presence of thermal gradients \cite{Moretti2017, Schlickeiser2014}. When heated locally, a DW must move towards the hottest regions in order to minimize its free energy. Therefore, the temperature gradient acts as an effective field that drives the DW towards the regions with higher temperature if it is greater than the depinning field at the laser temperature \cite{Schlickeiser2014}. Consequently, here the question is whether a MCD in the order of 0.5\% - 2\% as reported in the literature \cite{Gorchon2016, Medapalli2016} for ferromagnetic materials would induce a sufficient thermal gradient to unpin the DW. One can calculate the effective magnetic field for 2\% of MCD using the same model as in Ref. \cite{Torrejon2012} considering, for Pt/Co/Pt thin films, a DW with a surface energy $\sigma$\textunderscript{S} of 8 mJ.m\textsuperscript{-2}, a saturation magnetization of \textit{M}\textunderscript{S} = 1720 emu.cm\textsuperscript{-3}, a Curie temperature of \textit{T}\textunderscript{C} = 650 K, an initial temperature \textit{T}\textunderscript{0} = 300 K and a thermal gradient of about 1 K.nm\textsuperscript{-1}. An estimated field of 7 mT is found, which is close to the value used in the previously discussed athermal model. Hence, MCD produces a field that can induce DWD. From the results in Fig. 1(b), to explain the direction of the DW motion based on MCD, $\sigma$\textsuperscript{+} (resp. $\sigma$\textsuperscript{-}) needs to be more absorbed by a \textit{M}\textsuperscript{-} (resp. \textit{M}\textsuperscript{+}) domain. In such case, pumping the DW with $\sigma$\textsuperscript{+} laser pulses, the temperature would locally be higher in the \textit{M}\textsuperscript{-} domain, which would lead to an expansion of the \textit{M}\textsuperscript{+} domain. However, to confirm the role of MCD, it is important to know the actual direction of the temperature gradient and which magnetization state is the least absorbent for both helicities.

\section{Conclusion}
In conclusion, we have demonstrated that it is possible to observe helicity-dependent laser-induced domain wall motion in Co/Pt multilayer thin films, which show all-optical helicity-dependent magnetization switching. The reported domain wall displacement could be achieved either with fs- or ps-laser pulses with a displacement of $\sim$ 2 nm per 2-ps pulse. In order to compare it to any other stimuli-based DW motion, DW inertia during the laser pulse has to be studied. Interestingly, it was discovered that the process of domain wall motion induced by light involves the balance of three contributions, the domain wall pinning, the temperature gradient across the DW due to DC laser heating and the effect of the helicity. The Fatuzzo-Labrune model successfully reproduced the experimental results of the DW displacement, making the Inverse Faraday Effect a likely explanation for the helicity effect on the DW, while the uncertainty about the direction of the MCD gradient raises doubt about its contribution. These findings provide valuable insights into the underlying mechanism of AO-HDS as it is showed that is intrinsically related to helicity-dependent DW motion.

\newpage
\begin{acknowledgments}
We would like to thank Ray Descoteaux for technical assistance, as well as Pierre Vallobra for help with the DC magneton sputtering deposition of the samples. This work was supported partly by the French PIA project `Lorraine Universit\'e d'Excellence', reference ANR-15-IDEX-04-LUE and NRI, the Office of Naval Research MURI Program.
\end{acknowledgments}



\begin{thebibliography}{99}

\bibitem{ElHadri2017} M. S. El Hadri, M. Hehn, G. Malinowski, S. Mangin, J.Phys. D: Appl. Phys. 50, 133002 (2017)

\bibitem{Kirilyuk2010} A. Kirilyuk, A. V. Kimel, and T. Rasing, Rev. Mod. Phys. 82, 2731 (2010)

\bibitem{Stanciu2007} C. D. Stanciu, F. Hansteen, A. V. Kimel, A. Kirilyuk, A. Tsukamoto, A. Itoh, and T. Rasing, Phys. Rev. Lett. 99, 047601 (2007).

\bibitem{Mangin2014} S. Mangin, M. Gottwald, C.-H. Lambert, D. Steil, and V. Uhlir, L. Pang, M. Hehn, S. Alebrand, M. Cinchetti, G. Malinowski, Y. Fainman, M. Aeschlimann, and E. E. Fullerton, Nat. Mater. 13, 286 (2014).

\bibitem{Schubert2014} C. Schubert, A. Hassdenteufel, P. Matthes, J. Schmidt, M. Helm, R. Bratschitsch, and M. Albrecht, Appl. Phys. Lett. 104, 082406 (2014).

\bibitem{Lambert2014} C. H. Lambert, S. Mangin, B. S. D. C. S. Varaprasad, Y. K. Takahashi, M. Hehn, M. Cinchetti, G. Malinowski, K. Hono, Y. Fainman, M. Aeschlimann, and E. E. Fullerton, Science 345, 1337 (2014).

\bibitem{Takahashi2016} Y. K. Takahashi, R. Medapalli, S. Kasai, J. Wang, K. Ishioka, S. H. Wee, O. Hellwig, K. Hono, and E. E. Fullerton, Phys. Rev. Applied 6, 054004 (2016).

\bibitem{Radu2011} I. Radu, K. Vahaplar, C. Stamm, T. Kachel, N. Pontius, H. A. Durr, T. A. Ostler, J. Barker, R. F. Evans, R. W. Chantrell, A. Tsukamoto, A. Itoh, A. Kirilyuk, T. Rasing, and A. V. Kimel, Nature 472, 205208 (2011).

\bibitem{Ostler2012} T. A. Ostler, J. Barker, R. F. L. Evans, R. W. Chantrell, U. Atxitia, O. Chubykalo-Fesenko, S. El Moussaoui, L. Le Guyader, E. Mengotti, L. J. Heyderman, F. Nolting, A. Tsukamoto, A. Itoh, D. Afanasiev, B. A. Ivanov, A. M. Kalashnikova, K. Vahaplar, J. Mentink, A. Kirilyuk, T. Rasing, and A. V. Kimel, Nat. Commun. 3, 666 (2012).

\bibitem{Khorsand2012} A. R. Khorsand, M. Savoini, A. Kirilyuk, A. V. Kimel, A. Tsukamoto, A. Itoh, and T. Rasing, Phys. Rev. Lett. 108, 127205 (2012).

\bibitem{Cornelissen2016} T. D. Cornelissen, R. C\'ordoba, and B. Koopmans, Appl. Phys. Lett. 108, 142405 (2016).

\bibitem{Gorchon2016} J. Gorchon, Y. Yang, and J. Bokor, Phys. Rev. B 94, 020409 (2016)

\bibitem{ElHadri2016a} M. S. El Hadri, P. Pirro, C.-H. Lambert, S. Petit-Watelot, Y. Quessab, M. Hehn, F. Montaigne, G. Malinowski, and S. Mangin, Phys. Rev. B 94, 064412 (2016).

\bibitem{Medapalli2016} R. Medapalli, D. Afanasiev, D. Kim, Y. Quessab, S. A. Monotoya, A. Kirilyuk, T. Rasing, Alexey V. Kimel, and E. E. Fullerton, arXiv:1607.02505.

\bibitem{Parkin2008} S. S. P. Parkin, M. Hayashi, and L. Thomas, Science 320, 190-194 (2008). 

\bibitem{Hayashi2008} M. Hayashi, L. Thomas, R. Moriya, C. Rettner, and S. S. P. Parkin, Science 320, 209-211 (2008). 

\bibitem{Allwood2005} D. A. Allwood, G. Xiong, C. C. Faulkner, D. Atkinson, D. Petit, R. P. Cowburn, Science 309,1688-1692 (2005).

\bibitem{Metaxas2007} P. J. Metaxas, J. P. Jamet, A. Mougin, M. Cormier, J. Ferre, V. Baltz, B. Rodmacq, B. Dieny, and R. L. Stamp, Phys. Rev. Lett. 99, 217208 (2007).


\bibitem{Malinowski2011} G. Malinowski, O. Boulle and M. Kl\"aui, J. Phys. D: Appl. Phys. 44, 384005 (2011).

\bibitem{Miron2011} I. M. Miron, T. Moore, H. Szambolics, L. D. Buda-Prejbeanu, S. Auffret, B. Rodmacq, S. Pizzini, J. Vogel, M. Bonfim, A. Schuhl, and G. Gaudin, Nat. Mater. 10, 419-423 (2011).

\bibitem{Lahtinen2012} T. H. E. Lahtinen, K. J. A. Franke, and S. Van Dijken, Scientific Reports 2, 258 (2012).

\bibitem{Shepley2015} P. M. Shepley, A. W. Rushforth, M. Wang, G. Burnell, and T. A. Moore, Scientific Reports 5, 7921 (2015).

\bibitem{Torrejon2012} J. Torrejon, G. Malinowski, M. Pelloux, R. Weil, A. Thiaville, J. Curiale, D. Lacour, F. Montaigne, and M. Hehn, Phys. Rev. Lett. 109, 106601 (2012).

\bibitem{Tetienne2014} J.-P. Tetienne, T. Hingant, J.-V. Kim, L. Herrera Diez, J.-P. Adam, K. Garcia, J.-F. Roch, S. Rohart, A. Thiaville, D. Ravelosona, V. Jacques, Science, 344, 1366-1369 (2014).

\bibitem{Moretti2017} S. Moretti, V. Raposo, E. Martinez, and L. Lopez-Diaz, Phys. Rev. B 95, 064419 (2017).

\bibitem{Schlickeiser2014} F. Schlickeiser, U. Ritzmann, D. Hinzke, and U. Nowak, Phys. Rev. Lett. 113, 097201 (2014).

\bibitem{Bersweiler2016} M. Bersweiler, K. Dumesnil, D. Lacour and M. Hehn, J. Phys.: Condens. Matter 28, 336005 (2016).

\bibitem{ElHadri2016b} M. S. El Hadri, P. Pirro, C.-H. Lambert, N. Bergeard, S. Petit- Watelot, M. Hehn, G. Malinowski, F. Montaigne, Y. Quessab, R. Medapalli, E. E. Fullerton, and S. Mangin, Appl. Phys. Lett. 108, 092405 (2016). 

\bibitem{Fatuzzo1962} E. Fatuzzo, Phys. Rev., 127, 1999 (1962).

\bibitem{Labrune1989} M. Labrune, S. Andrieu, F. Rio, and P. Bernstein, J. Magn. Mater., vol. 80, p. 211, (1989).



\end{thebibliography}
\end{document}